\documentclass{appolb}

\newcommand{\be}{\begin{equation}}
\newcommand{\ee}{\end{equation}}
\newcommand{\beeq}{\begin{eqnarray}}
\newcommand{\eeeq}{\end{eqnarray}}

\def\kbo{{\bf k}}

\def\lbo{{\bf l}}

\def\rbo{{\bf r}}
\def\bbo{{\bf b}}
\def\xbo{{\bf x}}
\def\ybo{{\bf y}}

\usepackage{epsfig}
\begin{document}
\title{Saturation effects in DIS at low $x$%
\thanks{Plenary presentation at the X International Workshop
        on Deep Inelastic Scattering (DIS2002), Cracow, 30 April -- 4 May 2002.
%        \\
%        Supported by the Polish KBN grant No. 5 P03B 144 20 and by
%        the  Deutsche Forschungsgemeinschaft.
        }
}
\author{K. Golec-Biernat
\address{II Institute of Theoretical Physics, University of Hamburg, Hamburg,  Germany\\
{\it and} \\ H. Niewodnicza\'nski
Institute of Nuclear Physics, Krak\'ow, Poland}
}
\maketitle
\begin{abstract}
We review  basic ideas related to saturation of parton densities
at small value of the Bjorken variable $x$. A special emphasis is put
on interpreting the results from the $ep$ deep   inelastic scattering experiments
at HERA.
\end{abstract}

\section{Introduction}
The idea of parton saturation naturally arises in 
deep inelastic scattering  (DIS) at small values of the Bjorken variable $x$,
studied currently at the HERA collider in DESY.
The main measured small-$x$ effect is a strong rise of
the proton structure function $F_2$ in the limit $x\rightarrow 0$,
for fixed virtuality $Q^2$. Interpreted with the help of linear evolution
equations of  Quantum Chromodynamics,
this is a reflection of increasing parton densities (sea quarks
and gluons). Formally, the parton rise is so strong that for sufficiently small  $x$
the computed cross section violates unitarity,  which signals that
important physical  effects were neglected in the approximation leading 
to the linear evolution equations. These effects are responsible for
{\it saturation} of the parton densities by taming their strong rise.
Whether the saturation effects are already visible in the $F_2$ at HERA is
disputable, the linear DGLAP evolution equations describe the data for
$Q^2>2~\mbox{\rm GeV}^{2}$.
However, more exclusive diffractive processes strongly hint towards the affirmative
answer to the posed question. Moreover, the transition region of
$Q^2\simeq 1~\mbox{\rm GeV}^{2}$
for $F_2$, where the DGLAP analysis encounters basic difficulties, is easily described
within an approach based on parton saturation. An important result of saturation
is the existence of a saturation scale which is reflected in a new scaling law
for inclusive DIS cross section. The small-$x$ data do really show such a regularity.

In the following we present basic concepts leading to the notion of parton saturation
and discuss the relevance of this effect for experimental data, mainly from HERA.

%%%%%%%%%%%%%%%%%%%%%%%%%%%%%%%%%%%%%%%%%%%%%%%%%%%%%%%%%%%%%
\section{Linear evolution equations}

In the standard description of $ep$ DIS, the structure function
$F_2$ is determined by the quark and antiquark densities in the proton,
\be
\label{eq:1}
F_2(x,Q^2)\,=\,{ \sum_f}\,\, e_f^2\,x\,
\{q_f(x,Q^2)+{\bar{q}}_f(x,Q^2)\}\,+\,{\cal{O}}({\alpha}_s)\,,
\ee
where the summation is done over quark flavours. As usual, $x=Q^2/(2 P\cdot q)$ and
$Q^2=-q^2$, where $P$ and $q$ are the proton and  virtual photon momenta.  The quark
densities together with the gluon density $g(x,Q^2)$
satisfies the DGLAP evolution equations \cite{DGLAP}.
In  the  matrix notation, ${\bf{q}}=(q_f,\bar{q}_f,g)$,
\be
\label{eq:2}
\frac{\partial {\bf{q}}(x,Q^2)}{\partial \ln Q^2}
\,=\,
\int_x^1 \frac{dz}{z}\ {\bf{P}}(x/z,Q^2)\ {\bf{q}}(z,Q^2)\,
\ee
where the matrix of splitting functions is  computed
perturbatively,
\be
\label{eq:2a}
{\bf{P}}(z,Q^2)\,=\,\alpha_s(Q^2)\ {\bf{P}}^{(1)}(z)\,+\,\alpha_s^2(Q^2)\
{\bf{P}}^{(2)}(z)\,+ \,...\,.
\ee
Eq.~(\ref{eq:2}) allows to determine the scale dependence
in (\ref{eq:1}) provided initial conditions at some scale $Q_0^2$ are given.
This is done by fitting their form in $x$ to the existing DIS data. In this way
logarithmic scaling violation of $F_2$ is explained. The above description
applies to $Q^2\gg \Lambda^2_{QCD}$ for perturbative QCD to be valid.
How low in
$Q^2$ and in $x$ the pure DGLAP approach is applicable
should be determined from the analysis
of data. In principle, the values of $x$ and $Q^2$ indicating that
boundary are correlated, i.e. $Q^2=Q^2(x)$.

\begin{figure}[t]
  \vspace*{0.0cm}
     \centerline{
           \epsfig{figure=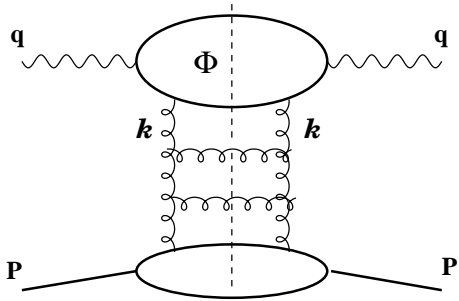,width=6cm}
           }
\vspace*{0.5cm}
\caption{\it $k_\perp$-factorization in DIS, eq.~(\ref{eq:4}).
\label{fig:1}}
\end{figure}

The low $Q^2$ region brings an issue of the $(1/Q^2)^n$ corrections (twist expansion)
to $F_2$, treated systematically in the
operator product expansion of two electromagnetic currents. Eq.~(\ref{eq:1})
takes into account only the logarithmic  contribution in this expansion.  For moderate $x$
this is a dominant contribution, as shown by successful DGLAP analyses of DIS data.
For $x\ll 1$, however,  each term in the twist expansion,
\be
\label{eq:3}
F_2(x,Q^2)\,=\,F_2^{(0)}(x,\ln Q^2)
\,+\,F_2^{(1)}(x,\ln Q^2)\,\frac{M^2}{Q^2}\,+\,...\,,
\ee
is equally important due to the presence of large logarithms $\ln(1/x)$.
Thus, a new systematics is necessary allowing to resum these large logarithms
in the small-$x$ limit, independent of the twist expansion.
This is done through   the $k_\perp$-factorization formula \cite{KTFAC}
which contains all twists
\be
\label{eq:4}
F_2(x,Q^2)\,=\,\int \frac{d^2 \kbo}{\kbo^4}\, \Phi(Q^2,\kbo)   \,f(x,\kbo)
\ee
where $\Phi(Q^2,\kbo)$ is the virtual photon impact factor describing the process
$\gamma^*\rightarrow q\bar{q}\rightarrow  \gamma^*$, see Fig.~\ref{fig:1}.
The function $f(x,\kbo)$
is the unintegrated gluon distribution, related to the gluon density
at  large $Q^2$ and small $x$  by
\be
\label{eq:5}
x g(x,Q^2)\,=\,\int \frac{d^2 \kbo}{\pi \kbo^2}\,f(x,\kbo)\,\Theta(Q^2-k^2).
\ee
In the above $\kbo$ is a two-dimensional vector of transverse momenta of two gluons which couple
to the quark loop  in four possible ways.
Notice the integration over all values of  $|\kbo|$ in (\ref{eq:4}).
This means that the region of small momenta needs special attention to avoid problems
with perturbative stability.
The  unintegrated gluon distribution satisfies the BFKL equation \cite{BFKL}
which can be written as an evolution equation in the rapidity $Y=\ln(1/x)$.
In the leading order approximation, when the terms proportional to
$(\alpha_s\ln(1/x))^n$ are resummed,
\be
\label{eq:6}
\frac{\partial f(x,\kbo) }{\partial Y}\,=\,
%(K\otimes f)(x,\kbo)\equiv
\frac{3\alpha_s}{\pi}\!\!
\int\! \frac{d^2 \lbo}{\pi \lbo^2} \left\{
f(x,\kbo+\lbo)\,-\,\Theta(\kbo^2-\lbo^2)\,f(x,\kbo)
\right\}.
\ee
Diagrammatically, the BFKL equation resums
the ladder diagrams with   two exchanged  (reggeized) gluons  interacting through
the real gluons in the rungs strongly ordered in rapidity. Each emitted gluon leads to
the large logarithm $\alpha_s \ln(1/x)$.
Since this is a colourless exchange which gives
the dominant behaviour at large energy $\sqrt{s}$ of the $\gamma^*p$ system
$(x=Q^2/s)$,  it is termed the BFKL pomeron.

\begin{figure}[t]
  \vspace*{0.0cm}
     \centerline{
           \epsfig{figure=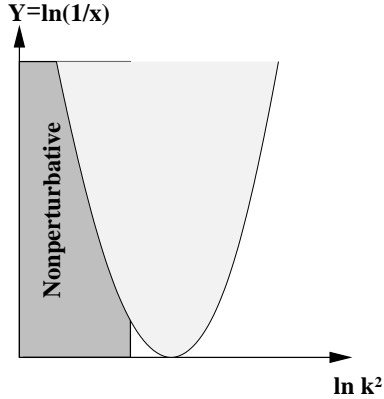,width=5cm}
           }
\vspace*{-0.0cm}
\caption{\it Diffusion pattern from the BFKL equation.
\label{fig:2}}
\end{figure}

The solution of eq.~(\ref{eq:6}) gives the $x$ dependence of $F_2$ at
small $x$. In the limit $x\rightarrow 0$
and $Q^2$ fixed, it is given by the  azimuthally symmetric
saddle point solution
\be
\label{eq:7}
\frac{f(x,k^2)}{\sqrt{k^2}}\,=\, x^{-12\alpha_s \ln 2/\pi}\,\,
\frac{\exp(-\ln^2(k^2/k_0^2)/D)}{\sqrt{\pi D}}
\ee
where $D\sim \ln(1/x)$. Thus for decreasing $x$, the solution features
a power-like  rise  with $x$ and {\it diffusion} in $\ln k^2 $.
Even though at some initial $x=x_0$ the solution is concentrated around
$k^2\approx k_0^2$, it diffuses into the region of larger and smaller
values of $k^2$ for $x\rightarrow 0$, see Fig.~\ref{fig:2}.
Diffusion into the infrared region means that the
BFKL approximation is strongly sensitive to nonperturbative effects since
the cross section might be dominated by a contribution
from $|\kbo|\simeq\Lambda$.
This problem can be phenomenologically cured by a separate
treatment of the infrared domain \cite{KMS}. However, infrared diffusion signals
fundamental difficulty which reflects incompleteness of the BFKL
approximation. This is intimately related to  violation of unitarity
reflected in the Froissart bound: $F_2 \le c \ln^2(1/x)$,
in contrast to the power-like rise in (\ref{eq:7}). The next-to-leading corrections
to the BFKL equation, see \cite{NLLBFKL} and references therein,
weaken the rise but the two discussed problems remain.

The missing physical effect in the BFKL approximation is parton saturation.
In the infinite momentum frame of the proton,  the following picture
of this phenomenon exists. The transverse size of gluons
with transverse momentum $\kbo$ is proportional to $1/|\kbo|$.
For large $|\kbo|$, the BFKL mechanism
of gluon radiation, $g\rightarrow gg$, populates the transverse space with
large number  (per unit of rapidity) of small size gluons.
The same mechanism also applies to large size gluons with small transverse momenta.
In this case, however, the BFKL approach is incomplete since large gluons strongly
overlap and fusion processes, $gg\rightarrow g$, are equally important.
By taking these processes into account unitarity is restored
and infrared diffusion cured. The key element for this is the emergence
of a rising with rapidity (energy) {\it saturation scale}  $Q_s(Y)$, which is a fundamental
property of parton saturation \cite{GLR,SATSCALE}.
The density of large gluons  with $|\kbo|<Q_s(Y)$
no longer strongly rises and unitarity is restored.
The low momentum contribution to the cross section
is dominated by $|\kbo|\simeq Q_s(Y)\gg  \Lambda_{QCD}$ for sufficiently large $Y$,
which allows to avoid
the problem of strong sensitivity to the infrared domain. In the forthcoming
we show how this idea is realized in practice. For complementary introductions,
see \cite{SATREV}.

%%%%%%%%%%%%%%%%%%%%%%%%%%%%%%%%%%%%%%%%%%%%%%%%%%%%%%%%%%%%%%%%%%%%%%%%%%%%%%%%%%%%%%%
\section{The dipole picture}

\begin{figure}[t]
  \vspace*{-0.5cm}
     \centerline{
           \epsfig{figure=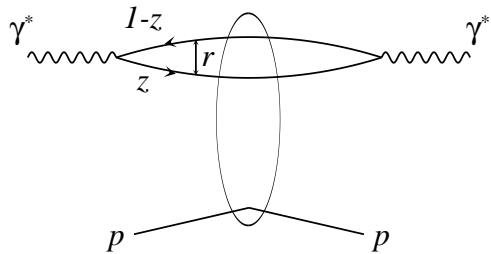,width=8cm}
           }
\vspace*{-0.0cm}
\caption{\it The illustration of factorization in eq.~(\ref{eq:8}).
\label{fig:3}}
\end{figure}

The problem of restoration of unitarity in the description of DIS at small $x$ was pioneered
by Gribov, Levin and Ryskin
in \cite{GLR}. In there nonlinear corrections to the DGLAP evolution equations
were presented and subsequently rigorously derived in the double logarithmic approximation by
Mueller and Qiu
in \cite{MUQIU}. In this approximation, $Q^2$ is large and $x$ is small such that terms
proportional to $(\alpha_s \ln x \ln Q^2)^n \sim 1$  are resummed.  We want to avoid the
high $Q^2$ assumption and approach the problem
in the approximation in which the BFKL formulation of DIS was discussed.
To this end we transform the $k_\perp$-factorization formula (\ref{eq:4}) into the Fourier
conjugate representation where the transverse momentum $\kbo$ is traded for
its conjugate transverse separation $\rbo$. Thus the following formula is found, see
e.g. \cite{JEFF},
\be
\label{eq:8a}
F_2={Q^2}/{(4\pi^2\alpha_{em})} (\sigma_T+\sigma_L),
\ee
and
\be
\label{eq:8}
\sigma_{T,L}(x,Q^2)\,=\,\int d^2\rbo\ dz\ |\Psi_{T,L}(Q^2,\rbo,z)|^2\,\,\hat\sigma(\rbo,x)
\ee
where $\sigma_{T,L}$ are $\gamma^*p$ cross sections for the indicated polarizations.
Here
$\rbo$ is the $q\bar{q}$ dipole   transverse separation
and $z$ is the longitudinal momentum fraction
of the dipole quark/antiquark with respect to the momentum $q^\prime=q+xp$.
The phy\-si\-cal interpretation of factorization in (\ref{eq:8}) is provided in the proton
rest frame. The virtual photon $\gamma^*$ splits into a $q\bar{q}$ pair long before
the interaction with the proton since the formation time, $\tau_{q\bar{q}}\sim 1/(x M)$,
is very large in the small-$x$ limit. This process is described by the lowest Fock
state light-cone wave function of the photon, $\Psi_{T,L}$ \cite{BKS}.
The interaction with the proton
is encoded in the $q\bar{q}$ dipole cross section $\hat\sigma(\rbo,x)$.
\begin{figure}[t]
  \vspace*{0.0cm}
     \centerline{
           \epsfig{figure=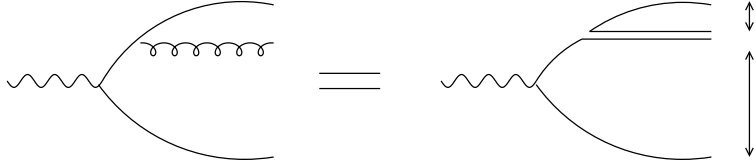,width=10cm}
           }
\vspace*{0.5cm}
\caption{\it Emission of a gluon in the large $N_c$ limit.
\label{fig:4}}
\end{figure}

The BFKL approximation has a nice formulation in the dipole
picture in  the large $N_c$ limit
\cite{MUDIPOL,NIKDIPOL}. In this limit, emission of a gluon by a quark
can be viewed as a creation of two dipoles out of the parent dipole, see Fig.~\ref{fig:4}.
Each new dipole can radiate a gluon with
much softer longitudinal momenta, which leads
to new dipoles. In this way the parent $q\bar{q}$ dipole $\rbo$ evolves into a collection
of dipoles $\rbo^\prime$ with the density $n(\rbo,\rbo^\prime,x)$.
The interaction of each dipole with the proton
occurs through exchange of  a single perturbative gluon. Thus in this
picture the dipole cross section equals
\be
\label{eq:9}
\hat\sigma(\rbo,x)\,=\,\int {d^2\rbo^\prime}\,
n(\rbo,\rbo^\prime,x)\,\sigma_{gp}(\rbo^\prime)\,
\ee
where $\sigma_{gp}(\rbo^\prime)$ describes the interaction of a dipole
$\rbo^\prime$ with the proton
The BFKL effects are located in evolution of the photon wave function,
the  large logarithms $\ln(1/x)$  come from strong ordering of
longitudinal momenta of the emitted gluons. Consequently,
the dipole density obeys the BFKL equation in the $\rbo$-space
\be
\label{eq:10}
\frac{\partial n(\rbo,\rbo^\prime,x) }{\partial Y}\,=\,
\frac{N_c \alpha_s}{\pi}\!\!
\int\! \frac{d^2 {\mbox{\boldmath $\rho$}}}{\pi {\mbox{\boldmath $\rho$}}^2} \left\{
n(\rbo+{\mbox{\boldmath $\rho$}},\rbo^\prime,x)\,-\,
\Theta(\rbo^2-{\mbox{\boldmath $\rho$}}^2)\,n(\rbo,\rbo^\prime,x)
\right\}.
\ee
The variable $\rbo^\prime$ is a parameter in  eq.~(\ref{eq:10}), thus the dipole cross section
(\ref{eq:9}) obeys the same equation. Notice that the integral kernel on the r.h.s. of
eq.~(\ref{eq:10}) has mathematically the same form as in the BFKL equation in the momentum space
(\ref{eq:6}).
Like in the momentum space, the
saddle point solution exhibits the power-like rise in $x$ and diffusion in
$\ln \rbo^2$:
\be
\label{eq:11}
\frac{n(\rbo,\rbo^\prime,x))}{\sqrt{r^2}}\,=\,x^{-4 N_c \alpha_s  \ln 2 /\pi}\,\,
\frac{\exp(-\ln^2(r^2/r^{\prime 2})/D)}{\sqrt{\pi D}}
\ee
and $D\sim \ln(1/x)$.
Now, infrared diffusion corresponds to an unlimited rise of the dipole
density for  large size parent dipoles.
From (\ref{eq:9}) and (\ref{eq:11}),   the BFKL dipole cross section
behaves as $x^{-\lambda}$  for any $r$
which leads to violation of unitarity when $x\rightarrow 0$ as in the momentum space.
Now, the saturation  effects would correspond  to interactions between dipoles.
Before discussing  QCD formulations of these effects, we present
a  phenomenological model which captures essential features of parton 
saturation and describes the DIS data.

%%%%%%%%%%%%%%%%%%%%%%%%%%%%%%%%%%%%%%%%%%%%%%%%%%%%%%%%%%%%%%%%
\section{The saturation model}

\begin{figure}[b]
  \vspace*{0.0cm}
     \centerline{
           \epsfig{figure=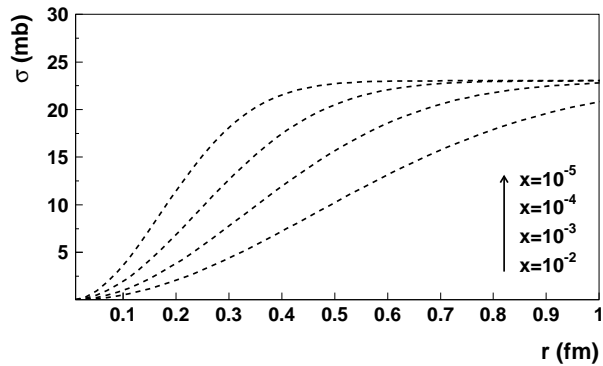,width=8cm}
           }
\vspace*{-0.0cm}
\caption{\it The dipole cross section in the saturation model.
\label{fig:5}}
\end{figure}

In this model \cite{GBW1}, the dipole cross section is bounded by an energy independent
value $\sigma_0$ which assures unitarity of $F_2$,
\be
\label{eq:12}
\hat{\sigma}(\rbo,x)\, =\, \sigma_0\,
\left\{1-\exp\left(-\frac{r^2}{4 R_0^2(x)}\right)\right\},
\ee
where  $R_0(x)$, called saturation radius,  is given by
\be
\label{eq:13}
R_0(x) = {1~\mbox{\rm GeV}^{-1}}\,\left({x}/{x_0}\right)^{\lambda/2}.
\ee
The three parameters, $\sigma_0=23~\mbox{\rm mb}$,
$\lambda\simeq 0.3$ and $x_0=3\cdot 10^{-4}$, were fitted to all small-$x$ DIS data
with $x<10^{-2}$.
The resulting curves for different values of $x$ are shown in Fig.~\ref{fig:5}.
At small $r$,  $\hat{\sigma}$ features colour transparency,
$\hat\sigma\sim r^2$, which is perturbative QCD phenomenon, while
for  large $r$, $\hat\sigma$ saturates, $\hat\sigma\simeq\sigma_0$.
The transition between the two regimes is governed by the increasing with $x$ saturation
radius $R_0(x)$. Therefore, an important feature of the model is that {\it for decreasing $x$,
the dipole cross section saturates for  smaller dipole sizes}, see
Fig.~\ref{fig:5}.  The BFKL rise with $x$
is encoded in the small $r$ part of  $\hat{\sigma}$ since for $r\ll R_0$,
$\hat{\sigma}\sim x^{-\lambda}$.
In contrast to the BFKL result, however, with decreasing $x$ the saturation radius
$R_0(x)$ gets smaller and consequently $\hat{\sigma}$ never exceeds ${\sigma_0}$.
This is the way infrared diffusion, which concerns large dipoles, is tamed
by the existence of the saturation scale:
\be
\label{eq:13a}
Q_s^2(x)\equiv 1/R_0^2(x)\,\sim\, x^{-\lambda}\,.
\ee
Notice that  the saturation scale rises when $x\rightarrow 0$.
With the dipole cross section (\ref{eq:12}) several remarkable features of DIS data
are described.

\begin{figure}[b]
 \vspace*{-0.5cm}
     \centerline{
           \epsfig{figure=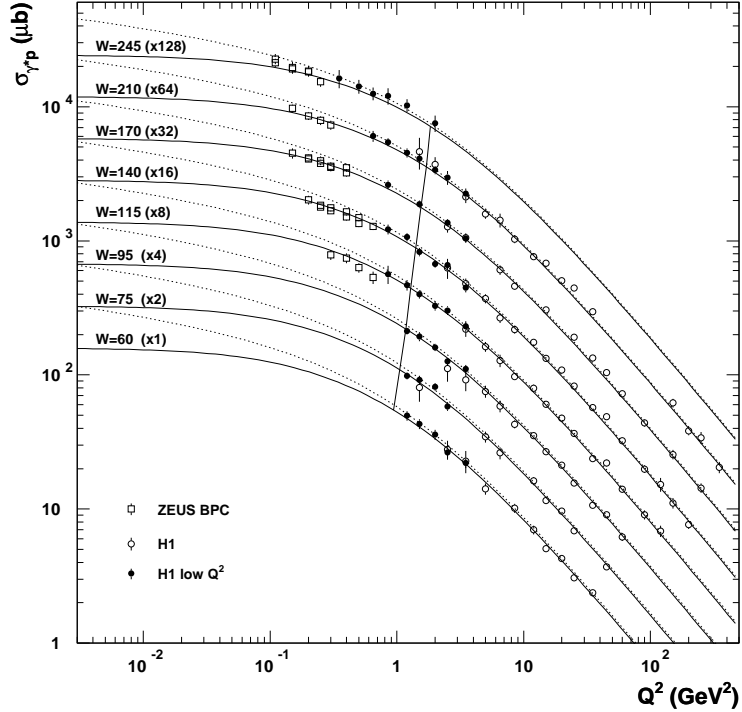,width=11cm}
           }
\vspace*{-0.0cm}
\caption{\it The cross section $\sigma_{\gamma* p}=\sigma_T+\sigma_L$
for fixed $\gamma^* p$ center-of-mass energy $W$ ($x=Q^2/W^2$).
The dotted lines,
for which $m_q=0$,  show the effect of the  dipole quark mass $m_q=140~\mbox{\rm MeV}$.
\label{fig:6}}
\end{figure}

\subsection{Transition to low $Q^2$}
First, the transition to low $Q^2$ values in the proton structure function.
This has an intuitive physical interpretation
if we relate the saturation radius to the mean transverse distance between partons.
If the partonic system is dilute and the $q\bar{q}$ dipole probe with a characteristic
size $1/Q$ is much smaller than the saturation radius, $1/Q\ll R_0(x)$,
the logarithmic behaviour is found \cite{GBW1,RING}
\be
\label{eq:14}
F_2\,\sim\, \frac{\sigma_0}{R_0^2(x)}\ln\left(Q^2 R_0^2(x)\right)\,.
\ee
In the opposite case, when $1/Q\gg R_0(x)$, the parton system looks dense for the probe
and
\be
\label{eq:15}
F_2\,\sim\, Q^2 \sigma_0 \ln\left(\frac{1}{Q^2 R_0^2(x)}\right)\,.
\ee
This transition is shown in Fig.~\ref{fig:6} for
$\sigma_{\gamma* p}=\sigma_T+\sigma_L\sim F_2/Q^2$.
With a nonzero quark mass in the $q\bar{q}$ dipole even the photoproduction data can be
described, which reflects a kind of continuity in the proposed description.
The transition region between the two regimes,
defined by the condition (critical line in the $(x,Q^2)$-plane)
\be
\label{eq:15a}
Q^2 R_0^2(x)\,=\,1\,, 
\ee
is found to be around $Q^2=1~\mbox{\rm GeV}^2$ at HERA kinematics (the solid line
across the model curves in Fig.~\ref{fig:6}). The critical line
also indicates the limit of validity of the twist expansion
(\ref{eq:3}) in the $(x,Q^2)$-plane  \cite{BGP}.
It is also interesting to analyze the energy dependence
resulting from the saturation model. If this is done for
increasing $Q^2$ values, smooth transition
between the soft $(F_2\sim x^{-0.08})$ and hard pomeron $(F_2\sim x^{-0.3})$
values is found \cite{GBW1}.

\begin{figure}[b]
  \vspace*{-0.5cm}
     \centerline{
           \epsfig{figure=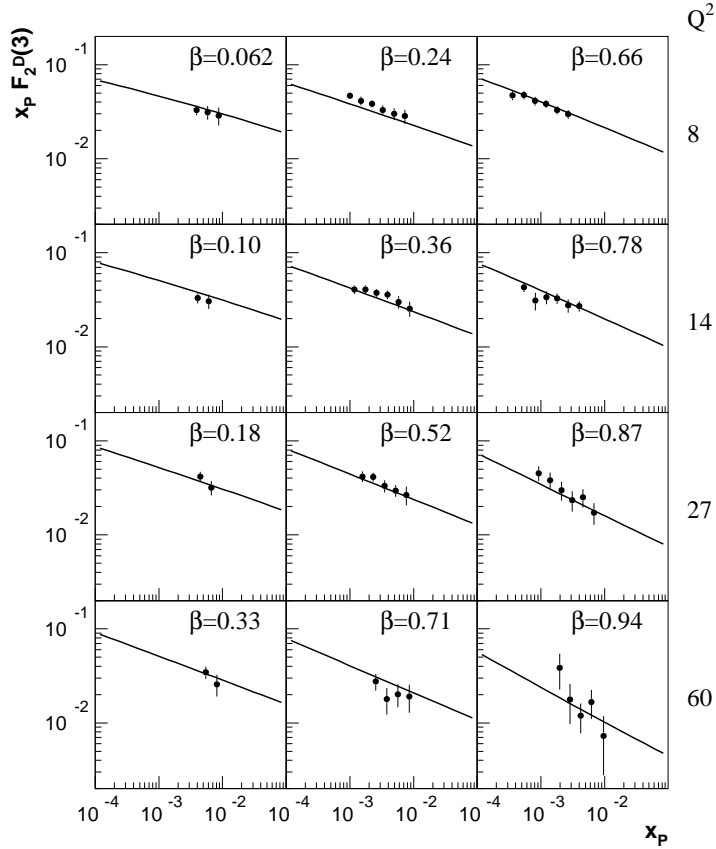,width=10cm}
           }
\vspace*{-0.0cm}
\caption{\it The diffractive data from ZEUS \cite{ZEUS} against the saturation model curves.
\label{fig:7}}
\end{figure}

\subsection{DIS diffraction}
Secondly, once the dipole cross section is determined from
the inclusive data analysis, it can be used as a prediction for diffractive
DIS \cite{GBW2} since for the $q\bar{q}$ pair in the diffractive final state
\be
\label{eq:15b}
\frac{d\,\sigma^D_{T,L}}{dt}_{\mid\, t=0}
\;=\;
\frac{1}{16\,\pi}\
\int d^2\rbo\, dz\
|\Psi_{T,L}(\rbo,z)|^2\ \hat\sigma^2(x,\rbo)\,.
\ee
The total diffractive  cross section $\sigma_{\gamma* p}^{D}=\sigma_T^D+\sigma_L^D$ is found
after dividing  (\ref{eq:15b}) by the diffractive slope $B_D$, taken from the experiment.
A very important  result of the saturation model is that with  the form (\ref{eq:12}),
the constant ratio as a function of $Q^2$ and $x$ of  the diffractive
and inclusive cross sections is naturally explained \cite{RING},
\be
\label{eq:16a}
\frac{\sigma_{\gamma* p}^D}{\sigma_{\gamma* p}}
\,\sim\, \frac{1}{\ln (Q^2 R_0^2(x))}\,.
\ee
With the diffractive $q\bar{q}$ and  $q\bar{q}g$ components,
the description of the diffractive data is quite satisfactory,
see Fig.~\ref{fig:7} \cite{GBW2}.

The diffractive interactions are
very important for tracing saturation effects since they are mainly sensitive
to intermediate dipole sizes, $r>2/Q$, which directly probe
the saturation part of the dipole cross section. In the inclusive
diffractive cross section, in contrast to the fully inclusive case, the
$r<2/Q$ contribution is suppressed by additional power of $1/Q^2$
\cite{GBW2,RING}.

\begin{figure}[b]
  \vspace*{-0.0cm}
     \centerline{
           \epsfig{figure=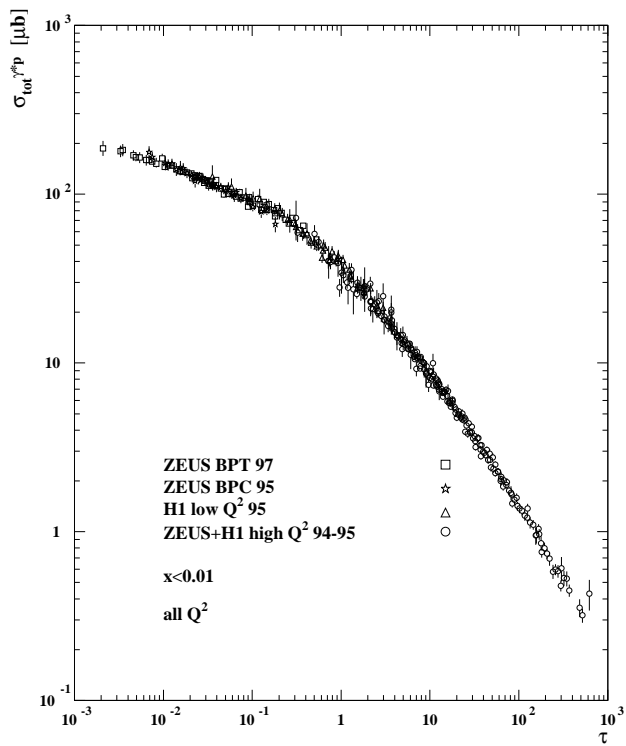,width=8.5cm}
           }
\vspace*{-0.0cm}
\caption{\it Geometric scaling for small-$x$ DIS data with $x< 10^{-2}$, the scaling variable
$\tau=Q^2R_0^2(x)$.
\label{fig:8}}
\end{figure}

\subsection{Geometric scaling}
Thirdly, the  particular feature of (\ref{eq:12}), i.e. its scaling form,
\be
\label{eq:17}
\hat\sigma(\rbo,x)\,=\,\hat\sigma\left(r/R_0(x)\right)\,,
\ee
leads to the
observation \cite{SGK} that
the virtual photon-proton cross section (or $F_2/Q^2$) is a function
of only  one dimensionless variable $\tau=Q^2 R_0^2(x)$  instead of $x$ and $Q^2$
separately
\be
\label{eq:18}
\sigma_{\gamma* p}(x,Q^2)\,=\,\sigma_{\gamma* p}(\tau)\,.
\ee
The new scaling (called geometric scaling),
valid in the small $x$ domain, is     shown  in Fig.~\ref{fig:8}
for the DIS data \cite{SGK}. The scaling variable $\tau$
is the ratio of two scales, the photon virtuality
$Q^2$ and the saturation scale (\ref{eq:13a}).
Therefore, in its essence geometric
scaling is a manifestation of the existence of the saturation scale.

It is interesting to look at the Regge limit, $x\rightarrow 0$ and
$Q^2$ fixed. From (\ref{eq:15}), the saturation model leads to
\be
\label{eq:19}
F_2\,\sim\,Q^2 \ln(1/x)
\ee
which is in accord with the Froissart bound, straightforwardly  applied to DIS:
$F_2\le c \ln^2(1/x)$. The logarithm in (\ref{eq:19})  comes from the wave function
$\Psi_T$ in eq.~(\ref{eq:8}), thus there is still room for additional logarithmic
dependence on $x$ of the dipole cross section.

%%%%%%%%%%%%%%%%%%%%%%%%%%%%%%%%%%%%%%%%%%%%%%%%%%%%%%%%%%%%%%%%%%%%%%%%%%%%%%%%%%%%%
\section{Nonlinear evolution equations}

The phenomenological success of the model (\ref{eq:12})
raises the question about its relation to QCD. In particular,
to what extent  is the transition to saturation justified by perturbative QCD?
The saturation scale $Q_s^2(x)\simeq 1~\mbox{\rm GeV}^2$ for $x\sim 10^{-4}$,
thus  we hope that at least the onset of saturation could be described
by weakly coupled QCD.

The problem of saturation effects was attacked by Kovchegov in the dipole picture
\cite{KOV}. He  found in the large $N_c$ limit  a nonlinear equation  which
is  a special case of more general hierarchy of equations derived by Balitsky \cite{BAL}.
Strictly speaking, the derived equation is only justified for a scattering on a large nucleus
since a certain class of diagrams, which could be relevant for the proton, is
suppressed for DIS on a nucleus. However, this equation can be considered as a QCD
model for $ep$ DIS at small $x$.
The basic quantity in  Kovchegov's formulation is the forward scattering amplitude
of the $q\bar{q}$ dipole (originated from the virtual photon)
on a nucleus, $N(\rbo, \bbo, Y)$. If $\xbo$ and $\ybo$ are the
transverse positions of the dipole quarks with respect to the center of the nucleus then
$\rbo=\xbo-\ybo$ and $\bbo=(\xbo+\ybo)/2$,
 where the latter quantity is the impact parameter
of the dipole.  The dipole cross section is given by
\be
\label{eq:20}
\hat\sigma(\rbo,x)\,=\,2 \int d^2 \bbo\, N(\rbo, \bbo, Y)\,.
\ee

\begin{figure}[t]
  \vspace*{-0.0cm}
     \centerline{
           \epsfig{figure=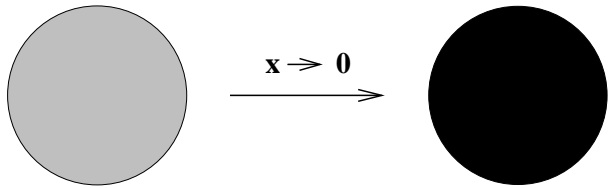,width=8cm}
           }
\vspace*{0.5cm}
\caption{\it Saturation (blackening) in the saturation model. Black means
$\hat\sigma=\sigma_0$.
\label{fig:9}}
\end{figure}

Comparing eq.~(\ref{eq:20}) with eq.~(\ref{eq:12}),
we see that in the saturation model the following
hypothesis on the form of $N$ is made
\be
\label{eq:21}
N(\rbo, \bbo, Y)\,=\,\left\{1-\exp\left(-\frac{r^2}{4 R_0^2(x)}\right)\right\}
\, \Theta(b-b_0)
\ee
where $\sigma_0=2\pi b_0^2$. The theta function can also be shifted into the argument of
the exponent. Physically, the saturation model corresponds to the proton being a disk
in the transverse plane with
a sharp boundary. Any edge effects are ignored. Saturation leads to a uniform blackening
of the disk with decreasing $x$ (for the parent dipole of a fixed size
$r$) without changing the disk size, see Fig.~\ref{fig:9}.

\begin{figure}[t]
  \vspace*{-0.0cm}
     \centerline{
           \epsfig{figure=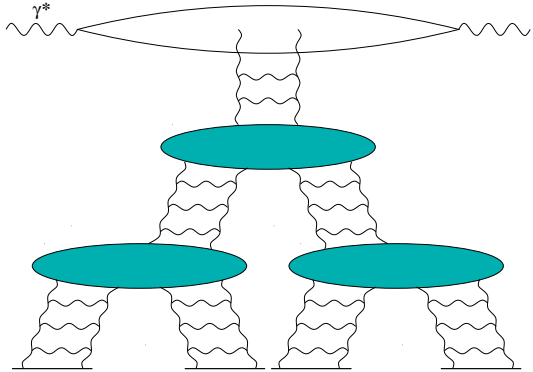,width=7cm}
           }
\vspace*{0.5cm}
\caption{\it The BFKL fan diagrams resummed by eq.~(\ref{eq:23}).
\label{fig:10}}
\end{figure}

In the approach of Kovchegov, the amplitude  $N(\rbo, \bbo, Y)$ is determined
dynamically  from the nonlinear evolution equation derived in the large $N_c$ limit.
Transforming back into the momentum representation,
\be
\label{eq:22}
\phi(\kbo,\bbo,Y)\,=\,\int \frac{d^2\rbo}{2\pi}\,\exp(-i\kbo\cdot\rbo)\,
\frac{N(\rbo,\bbo,Y)}{r^2}\,,
\ee
the following equation is found  for small dipoles $r\ll 1/\Lambda_{QCD}$
\be
\label{eq:23}
\frac{\partial \phi(\kbo,\bbo,Y)}{\partial Y}
\,=\,
(K\otimes \phi)(\kbo,\bbo,Y)\,-\,\overline{\alpha}_s\,\phi^2(\kbo,\bbo,Y)
\ee
where $\overline{\alpha}_s=N_c\alpha_s/\pi$ and $K$ is the kernel of the linear
BFKL equation  \cite{KOV}.
In the dipole picture in the leading $\ln(1/x)$ approximation,
the above equation describes not only the production
but also merging of dipoles in the dipole cascade.
In addition, an energy independent initial condition for the evolution
is supplied at $Y=Y_0$, which describes the scattering
of the parent $q\bar{q}$ dipole off the  nucleons through  multiple two-gluon
colour singlet exchange \cite{MUE90}. The effect of the nonlinear evolution
equation is the generation of
the interacting dipole cascade which brings the energy dependence of the solution.
In terms of  Feynman  diagrams, eq.~(\ref{eq:23}) resums fan diagrams with the BFKL
ladders and  triple pomeron couplings in the large $N_c$ limit
\cite{BRAUN1}, see Fig.~\ref{fig:10}.

\begin{figure}[t]
\vspace*{-0.0cm}
     \centerline{
           \epsfig{figure=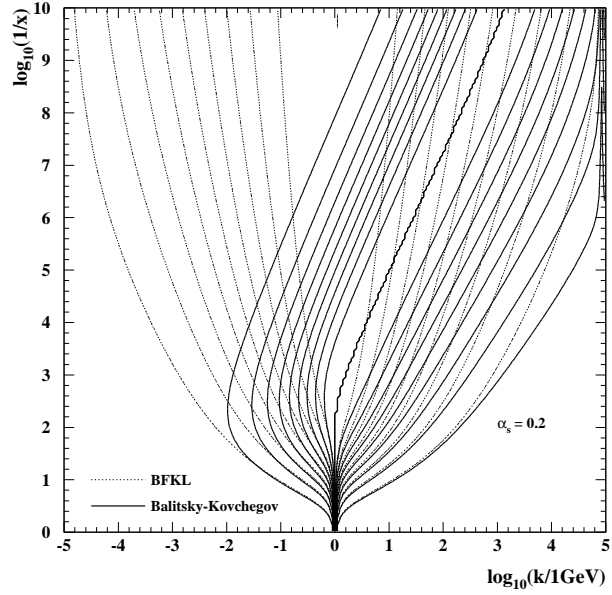,width=9cm}
           }
\vspace*{-0.8cm}
\caption{\it Diffusion in the BFKL and Balitsky-Kovchegov equations.
\label{fig:11}}
\end{figure}

Assuming the sharp disk model for the $b$-dependence,
both analytical and numerical studies of eq.~(\ref{eq:23})
\cite{KOV, BRAUN1,KOVLEV,KOVLUB, GMS} confirm the picture of blackening,
anticipated in the saturation model.
The dipole cross section (\ref{eq:20}) computed after solving
eq.~(\ref{eq:23}) features colour transparency for small $r$ and saturation at
large $r$ with a similar dependence on $x$, i.e. for smaller $x$ the saturation occurs for
smaller dipole sizes.
Moreover, the saturation scale emerges in such analyses. This is illustrated in detail
in  Fig.~\ref{fig:11} where the solution $\phi(k,Y)$ of eq.~(\ref{eq:23})  (solid lines)
and the solution of the corresponding
linear BFKL equation (dotted lines), projected on the $(k,Y)$-plane, are shown \cite{GMS}.
The BFKL
solution exhibits diffusion into small and large values of $k$ even though the initial
condition is concentrated at $k=k_0$. The nonlinear saturation effects
in the equation  (\ref{eq:23})
tame infrared diffusion, pulling the solution into the infrared
safe region.
The straight lines for the nonlinear solution reflect the scaling property:
$\phi(k,Y)=\phi(k/Q_s(Y))$, where the saturation scale $Q_s(Y)$ corresponds
to the central line $k=Q_s(Y)$, see \cite{GMS} for more details.  It is important to note that
the scaling sets in independent of initial conditions and is strictly valid for $k<Q_s(Y)$.
For  $k>Q_s(Y)$, however, the scaling is mildly violated. This was analyzed recently in
\cite{SCALVIOL} where the region of validity of geometric scaling was found.

Unfortunately, the agreement between the results based on the equation  (\ref{eq:23})
and the saturation model is
only qualitative and detailed studies  of the DIS data
with this equation are necessary.

\begin{figure}[b]
\vspace*{-0.0cm}
     \centerline{
           \epsfig{figure=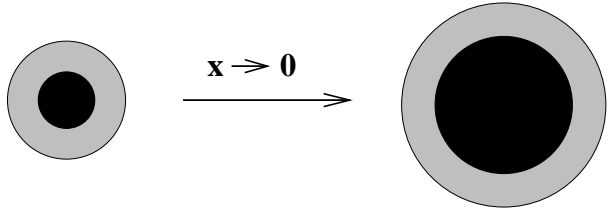,width=8cm}
           }
\vspace*{0.5cm}
\caption{\it Black disk expansion and saturation.
\label{fig:12}}
\end{figure}

An intriguing question, related to proton confinement, is the impact parameter dependence
of the dipole-proton forward scattering amplitude $N(\rbo,\bbo,Y)$.  Certainly,
the sharp disk picture with a fixed boundary is oversimplified if not questionable.
Recently, detailed
studies with eq.~(\ref{eq:23}) in the $\rbo$-representation were done
\cite{SCALVIOLB}, assuming
the initial condition $N(\rbo,\bbo,Y_0)=N(\rbo)\, S(\bbo)$ with some nonperturbative
profile function $S(\bbo)$ which falls as $exp(-2 m_{\pi}b)$ at large $b$.
The result of this analysis is shown in Fig.~\ref{fig:12}.
The black disk corresponds to saturation but now the saturation area
rises according to the  nonlinear evolution equation
(\ref{eq:23}) (or  the equation of the Colour Glass Condensate \cite{CGC})
when $x\rightarrow 0$.
Thus for each impact parameter $b$ unitarity limit is achieved. The grey area corresponds
to lower parton densities where the linear BFKL equation applies. Within this approach
the Froissart bound is saturated: $F_2 \sim \ln^2(1/x)$.

\begin{figure}[t]
\vspace*{-0.0cm}
     \centerline{
           \epsfig{figure=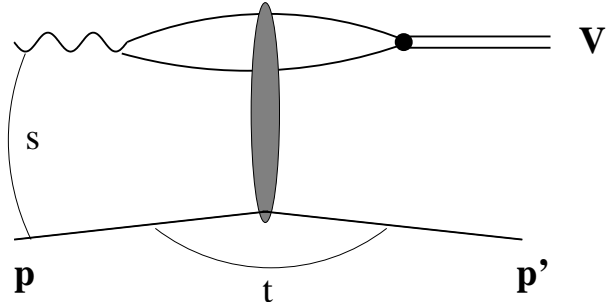,width=8cm}
           }
\vspace*{0.5cm}
\caption{\it Diffractive vector meson production.
\label{fig:13}}
\end{figure}

The impact parameter dependence of the forward scattering amplitude can be studied
through the $t$-dependence of diffractive vector meson production in DIS
\cite{MSM}, see Fig.~\ref{fig:13}. In this model, the vector meson production is
a three step process, the virtual photon splits into the $q\bar{q}$ pair which interacts
with the proton with the amplitude $A^{q\bar{q}-p}_{el}$, and then forms a meson described by
the wave function $\Psi_V$,
\be
\label{eq:24}
\frac{d\sigma_V}{dt~~}\,=\,\frac{1}{16\pi}\,\,\left|\int d^2\rbo\ dz\,
\overline{\Psi}_{\gamma*}(Q^2,\rbo,z)\,
A^{q\bar{q}-p}_{el}(\rbo,{\mbox{\boldmath $\Delta$}},x)\,\Psi_V(M_V,\rbo,z)\right|^2
\ee
where  $M_V$ is the vector meson mass. The amplitude
\be
\label{eq:24a}
A^{q\bar{q}-p}_{el}(\rbo,{\mbox{\boldmath $\Delta$}},x)\,=\,
 2\ \int d^2\bbo \, N(\rbo, \bbo, x)\,\mbox{\rm e}^{i{\mbox{\bf b}}
 \cdot {\mbox{\boldmath $\Delta$}}}\,
\ee
and ${\mbox{\boldmath $\Delta$}}$ is a two dimensional vector
of transverse momentum transferred from the proton into the
$q\bar{q}$ system, $t=-{\mbox{\boldmath $\Delta$}}^2$.
The presented above results are confirmed qualitatively although
the proton is not fully black at the central impact parameter
$b=0$ in  HERA kinematics. We would like to stress that
diffractive vector meson production  in $\gamma^*p$ scattering
is potentially the best process to study saturation
effects in DIS since the transverse size of the $q\bar{q}$ pair forming a meson is controlled
by the vector meson mass
\be
\label{eq:25}
\overline{r}\,=\,1/\sqrt{M_V^2+Q^2}\,.
\ee
Thus we expect saturation effects to be more important for larger (lighter)
vector mesons. The first analysis in this direction, done in \cite{MARA} with the saturation
model, seems to confirm this observation.

Let us finish by emphasizing that the nonlinear equation (\ref{eq:23}) is obtained
in the closed form  due to the large $N_c$ limit assumption. In general, in the leading
$\ln(1/x)$ approximation, a hierarchy of nonlinear equations is found by Balitsky \cite{BAL}.
These equations, written in a closed form by Weigert \cite{WEIGERT},
are equivalent to  the Color Glass Condensate renormalization group equation
\cite{CGC}, see \cite{SUMMARY} for details and more references.
A similar equation to eq.~(\ref{eq:23}) was also derived for diffractive
DIS \cite{KOVDIF}. A different approach to unitarization at small $x$ relies on the summation
of interacting reggeized gluons \cite{BKP}, forming a compound colour singlet state.
The BFKL equation is obtained assuming two
exchanged reggeized gluons. Recently, an energy spectrum
of the scattering amplitude
in high energy asymptotics with up to eight reggeized gluons was found
in the large $N_c$ limit \cite{KOT}. For the discussion of
the relation between the approach based on multiple BFKL pomeron exchanges, leading
to eq.~(\ref{eq:23}), and the approach based on exchanged reggeized gluons, see \cite{KOVNC}.

%%%%%%%%%%%%%%%%%%%%%%%%%%%%%%%%%%%%%%%%%%%%%%%%%%%%%%%%%%%%%%%%%%%%
\section{Conclusions and outlook}

We presented basic concepts of parton saturation. This phenomenon naturally appears
in DIS at small $x$ which is characterized by a strong rise of parton densities
in the nucleon. However, the strong rise is tamed in accordance with unitarity
of the description by the interactions between partons (mostly gluons). As a result,
gluons form a coherent system characterized by the saturation scale, and the distribution
of gluons with the transverse momentum below this scale no longer strongly rises.
This effects are important not only in $ep$ DIS, but also in nucleus-nucleus
collisions studied  at RHIC \cite{RHIC}.

Formulated in the dipole picture, the saturation effects impose strong bound on
the form of the dipole-proton cross section for the dipole sizes bigger than the inverse
of the saturation scale. This feature was build in the  phenomenological model \cite{GBW1} of
the dipole cross section which successfully describes the inclusive and diffractive DIS data
from HERA. In addition, based on the existence of the saturation scale, the new scaling
(``geometric scaling'') was proposed and successfully confronted with the data.
The QCD based attempt to justify the main features of the phenomenological description,
which gives rise to a formulation of a nonlinear evolution equation, was
presented.  Based on the numerical solution to this equation we showed the main effect of
saturation which suppresses diffusion into small values of transverse gluon momenta,
present in the approach based on the linear BFKL equation. We also discussed  the relevance of
the impact parameter studies which touch the problem of proton confinement.
Experimentally, this is done through the diffractive
vector meson production.

From the theoretical side, the
future work will concentrate on detailed analysis of nonlinear equations
describing parton saturation. Refraining from the large $N_c$ limit,
there are two equivalent formulations, the Colour Glass Condensate
equation \cite{CGC} and the equation derived by Weigert \cite{WEIGERT}
which beautifully summarizes the infinite hierarchy of the equations obtained by
Balitsky \cite{BAL}.  It would be good to have full handle over these equations.
From the phenomenological side, more precise data awaited from HERA II will
allow to test the parton saturation formulations in great detail. For this purpose,
the phenomenological saturation model (\ref{eq:12}) was refined in order to include the DGLAP
gluon evolution, expected at small $r$ \cite{BGK}. All important features of the
saturation in the original model, however,  were retained.

The attractiveness  of the presented description of saturation effects  to large extent
relies on the dipole formulation (\ref{eq:8}), obtained in the leading $\ln(1/x)$ approximation.
How this picture changes when the next-to-leading order corrections
to the formula (\ref{eq:4}) are computed, which
introduce the $q\bar{q}g$  Fock state in the photon wave function,
is not known yet.
The computations are under way \cite{GIES}.

\vskip 1cm
\centerline{\bf Acknowledgement}

This paper was supported in part by the Polish KBN grant No. 5 P03B 144 20 and by
the  Deutsche Forschungsgemeinschaft.

%%%%%%%%%%%%%%%%%%%%%%%%%%%%%%%%%%%%%%%%%%%%%%%%%%%%%%%%%%%%%%%%%%%%%%%%%%%%%%%%%%%%%

\end{document}